\documentstyle[prc,aps,twocolumn,epsf]{revtex}
\begin{document}
\twocolumn[\hsize\textwidth\columnwidth\hsize\csname
@twocolumnfalse\endcsname

\draft
\title{At the edge of nuclear stability: nonlinear quantum amplifiers}
\author{Attila Cs\'ot\'o}
\address{Department of Atomic Physics, E\"otv\"os University,
P\'azm\'any P\'eter s\'et\'any 1/A, H--1117 Budapest, Hungary}
\date{\today}

\maketitle

\begin{abstract}
We show that nuclear states lying at the edge of stability may show 
enormously enhanced response to small perturbations. For example, a 
0.1\% change in the strength of the strong nucleon-nucleon interaction
can cause almost a hundred times bigger change in the resonance
energy of the $0^+_2$ state of $^{12}$C. 
\end{abstract}
\pacs{PACS number(s): 21.10.-k, 21.30.-x, 21.45.+v, 26.20.+f}
\ \\
]

\narrowtext

\section{Introduction}

Recently, the birth of radioactive nuclear beams has made it possible
to perform extensive studies of the structure and reactions of nuclei
lying at the edge of stability. It was found that these nuclei posses
some really unusual properties, such as, for example, halo structure
\cite{Hansen}, which were largely unknown in stable systems. Here we 
would like to show that some nuclear systems close to the point of 
stability may exhibit a hitherto unrecognized interesting phenomenon, 
which we call nonlinear quantum amplifying. As the system moves 
toward the edge of stability, its response to small perturbations can 
become hugely amplified. This effect may find interesting 
applications in halo nuclei or in astrophysical processes.

We consider nuclei which have strong 2-body or 3-body clustering
nature. This means that their wave functions contain 2- or 3-cluster
structures with large weight, therefore the most important degrees of 
freedom are the relative motions between the clusters. For instance,
the low-lying states of $^7$Li and $^{12}$C are known to have strong
$^4{\rm He}+{^3{\rm H}}$ and $^4{\rm He}+{^4{\rm He}}+{^4{\rm He}}$
cluster structures, respectively. In addition, we assume that these
systems are below or slightly above the threshold of the lowest
possible breakup channel. If the energy of such a state, relative to
the 2-body or 3-body breakup threshold, is close to zero then one may
call it a weakly coupled few-body system. It is either a bound state
or a long-lived resonance. In any case, the system is barely held
together by the residual interactions acting between the clusters. 

It is interesting to see how the residual interaction between the
clusters behaves as the system goes from a weakly-bound state through
the breakup point to a resonance state. While this happens, the size
of the system increases. If the interaction between the clusters is
very short-ranged, then one can imagine that the residual interaction
drops down to zero as fast as the binding energy itself. Then nothing
special happens. However, one can imagine situations where the
residual force only slowly decreases but does not vanish, while the 
binding energy becomes zero. If this happens, then the binding energy
becomes very sensitive to small perturbations, for example, in the
nucleon-nucleon (N-N) force. 

In order to see if such situations are really realized, we study three
nuclear systems: the deuteron, the ground state of $^7$Li, and the 
$0^+_2$ state of $^{12}$C.

\section{Results}

\begin{figure}
\centerline{\centerline{\epsfxsize 7.5cm \epsfbox{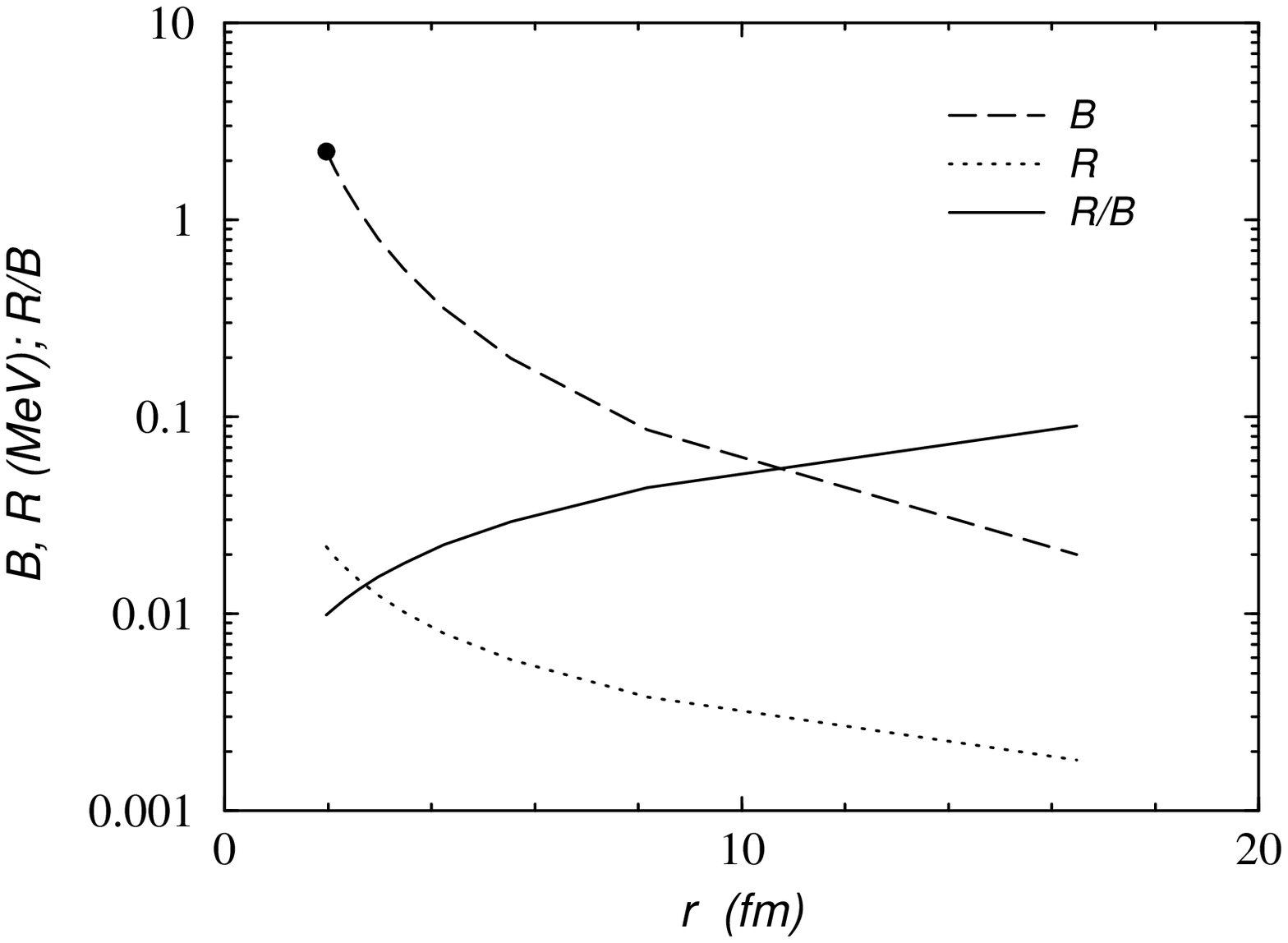}}}
\caption{The energy ($B=\vert E\vert$, where $E$ is the binding energy 
or resonance energy, relative to the breakup threshold), the response 
($R=\vert E_p-E_{p\times 1.001}\vert$, where $E_p$ and $E_{p\times
1.001}$ are the binding energies or resonance energies corresponding
to a given N-N force and another one which is stronger by 0.1\%,
respectively), and the $R/B$ ratio calculated for several artificial 
deuterons, as functions of the radius of the deuteron. The N-N 
interaction is chosen to be the Reid93 force \protect\cite{Reid93} 
in each case, with the strengths multiplied by a number $p$ (see the 
text). The black dot shows the real physical deuteron, given by our 
model.} 
\label{fig1}
\end{figure}

We first choose the simplest possible example, the deuteron. The 
bound-state problem of the deuteron is considered in the case of a 
modern realistic interaction, the Reid93 force of the Nijmegen group
\cite{Reid93}. The Schr\"odinger equation is solved by using the
method discussed in Ref.\ \cite{Payne}. In order to see how the
response of the deuteron to small perturbations in the force would
change if its binding energy were less than 2.22 MeV, we create
several artificial deuterons by changing the N-N force. We multiply
the strengths of each component of the N-N force (central, spin-orbit, 
and tensor) by $p(<1)$, and solve the Schr\"odinger equation
for each $p$ value. The binding energy as a function of the radius of
the deuteron is shown by the dashed line in Fig.\ \ref{fig1}. The last
deuteron which is generated has a 16.48 fm radius and 0.0201 MeV
binding energy. In this case, all strengths in the Reid93 force are
multiplied by $p=0.82$. For a weaker interaction, the deuteron becomes
unbound. 

For each artificial deuteron, we calculate its response to small
perturbations in the N-N force. Namely, we calculate the difference
between the energies of the deuteron corresponding to a $p$ value, 
and another one which comes from a 0.1\% stronger N-N force. We note,
that this response is closely related to the effective residual force
between the clusters (between the proton and the neutron in the case
of the deuteron). This response is shown by the dotted curve in 
Fig.\ \ref{fig1}. One can observe that the dotted curve goes down
somewhat more slowly than the dashed curve, as the radius increases.
As the $R/B$ ratio shows (solid line), close to the point of
instability the response can be rather large, compared to the binding
energy itself. The energy of a deuteron which is bound by only 20 keV
would change by 9\% in response to a 0.1\% change in the strength of
the N-N force.

We performed several test calculations using the deuteron model. In
heavier systems, discussed below, the use of realistic interactions is
not feasible in our model, and effective forces have to be considered. 
We checked that such forces produce virtually the same result as shown 
in Fig.\ \ref{fig1}. First we tested the Eikemeier-Hackenbroich (EH) 
interaction \cite{EH}, which gives a rather good overall description
of the $N+N$ scattering states and of the deuteron properties, but has
a Gaussian asymptotic behavior, instead of a Yukawa tail. Our second
effective interaction is even much simpler. The Minnesota (MN) force
\cite{MN} is designed to give the correct physical deuteron energy in
a $^3S_1$ model, without tensor coupling. Yet, even this extremely
simple interaction gives practically the same result as the EH or
Reid93 forces. We note that in the case of the EH interaction, the
deuteron becomes unbound at roughly the same $p\approx0.82$ strength
value as found in the case of the Reid93 force, while the MN
interaction needs a stronger reduction, $p\approx0.72$. The fact that
we get practically the same results for rather different forces means,
that the effect we see using effective forces is not an artifact
related to some shortcomings of those interactions. In the following
discussions we will use the MN force.

In heavier systems we cannot use the procedure of Ref.\ \cite{Payne}
to solve the Schr\"odinger equation. Instead, we use a variational
expansion of the relative-motion wave functions in terms of Gaussian
basis functions. We use this technique both for bound states and
narrow (low-energy) resonances. For the unbound states this procedure
is an approximation, which, however, should work rather well for
narrow states. In the case of $^7$Li, we checked that this is really
the case. In the case of the deuteron, the variational procedure gives
the same result as the method of Ref.\ \cite{Payne}. Of course in the
case of the very extended artificial deuteron with small binding
energy, our basis have to go out to very large radii. 

The second nucleus we study is $^7$Li at its ground state, which is 
known to have a strong two-body, $^4{\rm He}+{^3{\rm H}}$, clustering 
nature. Therefore, the most important degrees of freedom is the 
relative motion between $^4$He and $^3$H. In accordance with this 
fact, we describe $^7$Li by using a microscopic cluster model, 
similar to, e.g., Ref.\ \cite{li7}. The wave function of $^7$Li 
looks like 
\begin{equation}
\Psi^J={\cal A} \Bigl\{[\Phi^\alpha \Phi^t]_S\chi^{\alpha t}_L  
(\mbox{\boldmath $\rho$}) \Bigl\}, 
\label{li7}
\end{equation}
where ${\cal A}$ is the intercluster antisymmetrizer, $S=1/2$ is the 
intrinsic spin, $L=1$ is the relative orbital angular momentum, 
$J=3/2$ is the total angular momentum, the $\Phi^\alpha$ and $\Phi^t$ 
cluster internal states ($\alpha={^4{\rm He}}$ and $t={^3{\rm H}}$) 
are simple harmonic-oscillator shell-model functions, while $\chi$ is 
the wave function of the relative motion.  

\begin{figure}
\centerline{\centerline{\epsfxsize 7.5cm \epsfbox{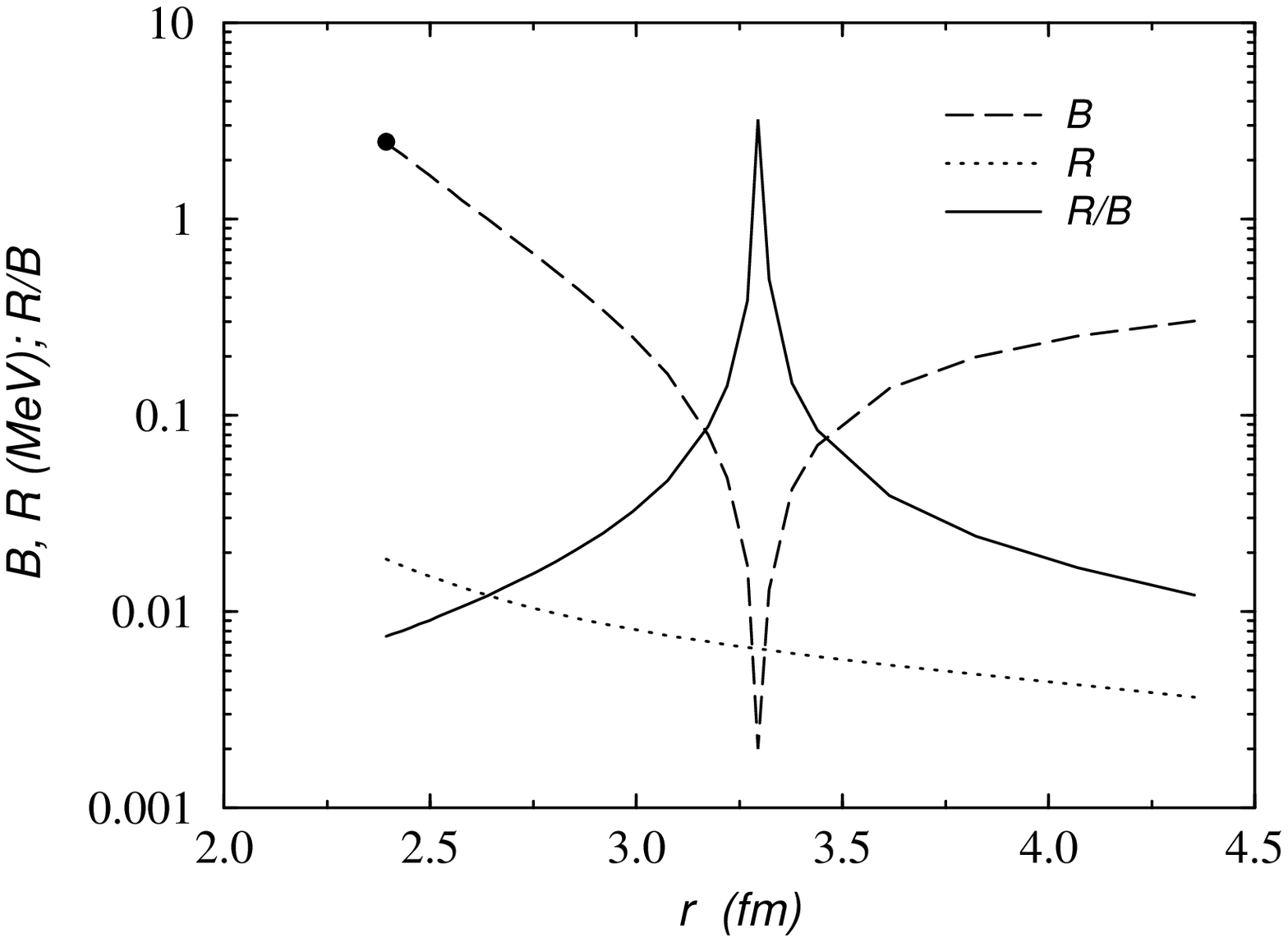}}}
\caption{The same as Fig.\ \protect\ref{fig1}, except for the ground
state of $^7$Li. The interaction is chosen to be the MN force 
\protect\cite{MN}.} 
\label{fig2}
\end{figure}

The results of our calculations for $^7$Li are shown in Fig.\ 
\ref{fig2}. One can observe a markedly different behavior of the 
binding energy as a function of the radius than it was in the case of 
the deuteron. As the $^4{\rm He}+{^3{\rm H}}$ two-body system is 
charged, it breaks up at a small radius, and the $^7$Li bound state 
quickly becomes a resonance. We note in passing, that we always 
modify only the strong forces, while leaving the Coulomb force 
intact. This way we can test the sensitivity of the system at the 
edge of stability to small perturbations in the strong coupling 
constant. 

One can see in Fig.\ \ref{fig2} that while the dashed curve drops to
zero very rapidly, the dotted curve of the response goes down rather
slowly. As a consequence, in the vicinity of the breakup point the
response to small perturbations in the N-N force gets hugely
amplified. 

We checked how well the pseudo-bound-state approximation
used by us works for the narrow $^7$Li resonances shown in Fig.\
\ref{fig2}. For this purpose, we localized these states as poles of
the scattering matrices, using the method discussed in Ref.\
\cite{pole}. As expected, the results coming from the bound-state 
approximation are all close to those coming from the correct resonance
description.

In both cases studied so far, the real physical system (shown by the
black dots on the figures) was relatively far from the point where the
large amplification phenomenon can appear. Our last example shows a
case where the real nuclear state lies close to the strong
amplification region. 

The $0^+_2$ state of $^{12}$C is situated only 380 keV above the
$3\alpha$-threshold, which makes this level one of the most important
nuclear states in astrophysics. Almost all carbon in the Universe is
synthesized through this state in red giant stars \cite{Rolfs}. We use
a three-cluster description of $^{12}$C, similar to that in Ref.\
\cite{c12}. The three-alpha wave function is given as
\begin{equation}
\Psi^J={\cal A} \Bigl\{ \Phi^\alpha \Phi^\alpha \Phi^\alpha\chi^{
\alpha(\alpha\alpha)}_{[l_1l_2]L} (\mbox{\boldmath $\rho
$}_1,\mbox{\boldmath $\rho$}_2) \Bigl\},
\label{c12}
\end{equation}
where $l_1=l_2=L=J=0$, and we concentrate on the second $0^+$ state.

\begin{figure}
\centerline{\centerline{\epsfxsize 7.5cm \epsfbox{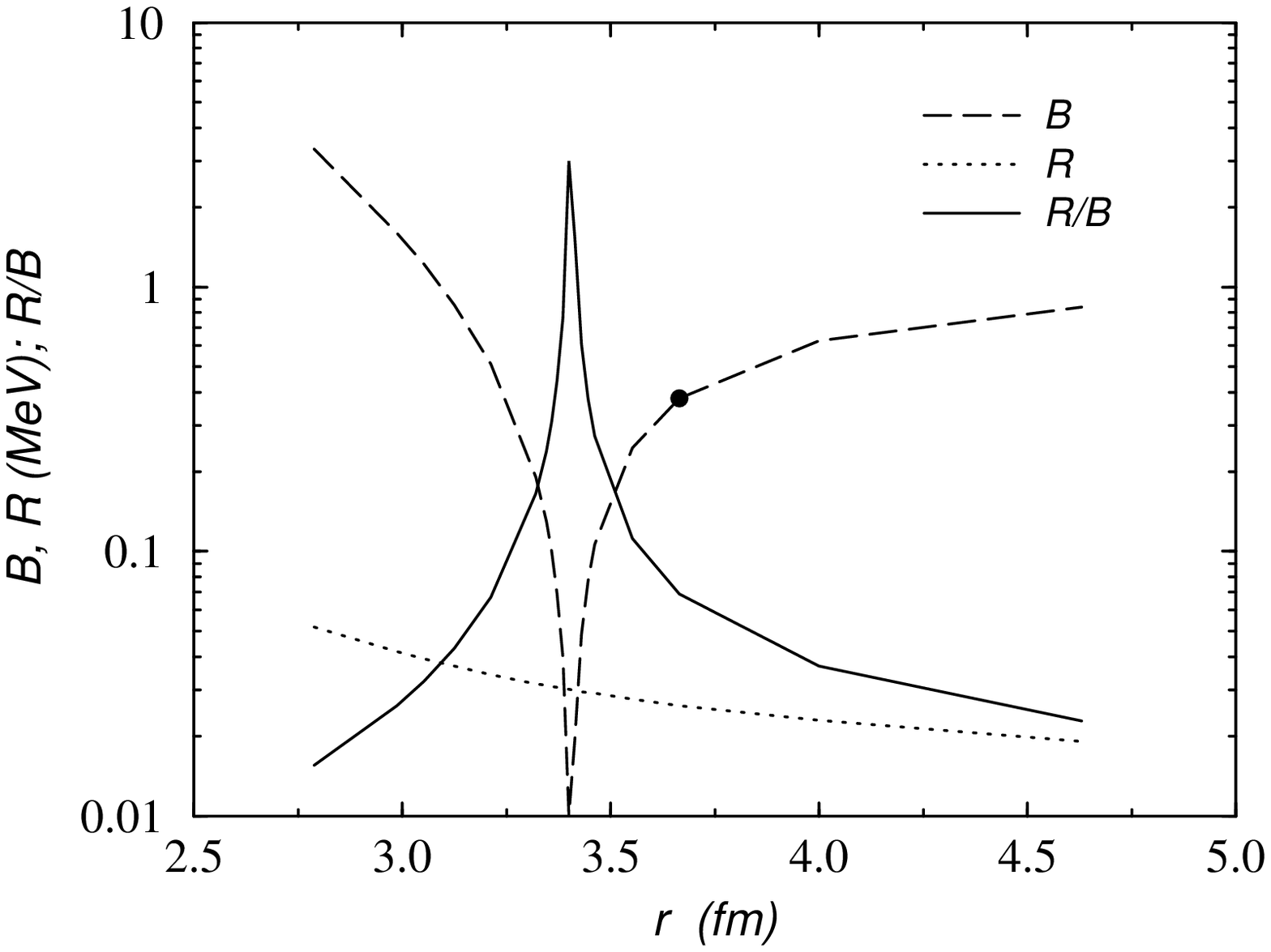}}}
\caption{The same as Fig.\ \protect\ref{fig2}, except for the $0^+_2$
state of $^{12}$C. Note that the vertical scale is different from 
those in Figs.\ \protect\ref{fig1} and \protect\ref{fig2}.} 
\label{fig3}
\end{figure}

We performed the same kind of calculations for this state as for the
deuteron and $^7$Li. The results are shown in Fig.\ \ref{fig3}. One can
see that the general behavior of the binding/resonance energy and the
response is qualitatively the same as in the case of $^7$Li. However,
this time the real physical position of the state is really close to
the interesting region. We find that a 0.1\% change in the strength of
the strong N-N force leads to a roughly 7\% change in the resonance 
energy of the physical $0^+_2$ state of $^{12}$C in our model. This 
sensitivity to the N-N force has a spectacular consequence in
astrophysical carbon synthesis. Careful studies of all the details of
the process show that a mere 0.5\% change in the strength of the N-N
force would lead to a Universe where there is virtually no carbon or
oxygen present \cite{carbon}. This makes carbon production one of the
most fine-tuned processes in astrophysics, leading to interesting
consequences for the possible values of some fundamental parameters of
the Standard Model \cite{Jeltema}.

\section{Conclusions}

We have shown that certain nuclear states lying at the edge of
stability may behave as nonlinear amplifiers: a tiny change in the N-N
interaction can get enormously amplified in the binding energy or
resonance energy. This behavior is mainly caused by the fact that the
residual interaction between the 2-3 clusters of the nucleus goes down
to zero more mildly than the binding energy itself. 

We can envision several possible applications of the nonlinear
amplification process discussed here. One application was exemplified
through the $0^+_2$ state of $^{12}$C. Because of the amplification 
phenomenon, some astrophysical processes can be very strongly 
fine-tuned. Halo nuclei are also natural candidates to show this 
phenomenon, as the halo effect itself is strongly connected to the 
disappearing binding energy.

In summary, we have presented an interesting property of some nuclear
systems lying at the edge of stability. Whether one can find any
useful applications of this feature, beyond the one example of the 
$0^+_2$ state of $^{12}$C, remains to be seen.

\acknowledgments

This work was supported by Grants from the OTKA Fund (D32513), 
the Education Ministry (FKFP-0242/2000) and the National Academy 
(BO/00520/98) of Hungary, and by the John Templeton Foundation 
(938-COS153). We are grateful to B.~F.\ Gibson, S.~A.\ Moszkowski, 
H.\ Oberhummer, and S.\ Weinberg for useful discussions in connection 
with the presented work, and G.~L.\ Payne for his kind help with his 
deuteron code.

%\onecolumn

\narrowtext

\narrowtext

\narrowtext

\end{document}